\documentclass[aps,twocolumn,superscriptaddress,nofootinbib,pre,floatfix]{revtex4-1}

\usepackage{url}

\usepackage[pdftex]{graphicx}
\usepackage{latexsym,amsmath,verbatim}
\usepackage{xcolor}

\usepackage[hidelinks,unicode=true]{hyperref}
\hypersetup{colorlinks=true,linkcolor=blue,urlcolor=blue,citecolor=blue,pdfhighlight=/N
}
\usepackage{lipsum}
\usepackage[english]{babel}
\usepackage{microtype}

\usepackage{bbold}
\usepackage[normalem]{ulem}

\newcommand{\av}[1]{\left\langle {#1} \right\rangle}

\newcommand{\be}{\begin{equation}}
\newcommand{\ee}{\end{equation}}

\graphicspath{{./figures/}}

\begin{document}
\title{The effect of delayed awareness and fatigue \\
on the efficacy of self-isolation in epidemic control}

\author{Giulia De Meijere}
\affiliation{Gran Sasso Science Institute, Viale  F.  Crispi  7,  67100  L’Aquila, Italy}
\affiliation{Istituto dei Sistemi Complessi (ISC-CNR), Via dei Taurini 19,
I-00185 Roma, Italy}

\author{Vittoria Colizza}
\affiliation{INSERM, Sorbonne Universit{\'e}, Pierre Louis Institute of Epidemiology and Public Health, Paris, France}
\affiliation{Tokyo Tech World Research Hub Initiative, Institute of Innovative Research, Tokyo Institute of Technology, Tokyo, Japan }
\author{Eugenio Valdano}
\affiliation{INSERM, Sorbonne Université, Pierre Louis Institute of Epidemiology and Public Health, Paris, France}

\author{Claudio Castellano}
\affiliation{Istituto dei Sistemi Complessi (ISC-CNR), Via dei Taurini 19,
I-00185 Roma, Italy}

\begin{abstract}
  The isolation of infectious individuals is a key measure of public health for the control of communicable diseases. However, 
  involving a strong perturbation of daily life, it often causes psychosocial distress, and severe financial and social costs. These may act as mechanisms limiting the adoption of the measure in the first place or the adherence throughout its full duration. In addition, difficulty of recognizing mild symptoms or lack of symptoms may impact awareness of the infection and further limit adoption. Here, we study an epidemic model on a network of contacts accounting for limited adherence and delayed awareness to self-isolation, along with fatigue causing overhasty
  termination. The model allows us to estimate the role of each ingredient
  and analyze the tradeoff between adherence and  duration of self-isolation. We find that the epidemic threshold is very sensitive to an effective compliance that combines the effects of imperfect adherence, delayed awareness and fatigue. If adherence improves for shorter quarantine periods, there exists an optimal duration of isolation, shorter than the infectious period.
  However, heterogeneities in the connectivity pattern, coupled to a reduced
  compliance for highly active individuals, may almost completely offset the effectiveness 
  of self-isolation measures on the control of the epidemic.
\end{abstract}

\maketitle

\section{Introduction}

A pillar of non-pharmaceutical interventions for the control of
COVID-19 pandemic is the isolation of individuals testing positive for
SARS-CoV-2 infection. The aim is to avoid onward propagation of the
disease, while contacts are traced to further break the chains of
transmission~\cite{Nature_review}. This measure, however, is met with
a set of challenges, as it has no immediate benefit for the index
case, but a number of downsides. It often causes psychosocial
distress~\cite{Brooks2020}, and it may have severe financial and social
costs impacting daily life, if a structured support program is not in
place.

Ideally, the measure should cover the entire duration of the
infectivity period. In practice, isolation may start when a person is
already infectious, typically at the onset of symptoms, or when
alerted by a contact tracing investigation. Also, the length of the infectious
period may be strongly variable across individuals~\cite{Sudre2020,who2020}. Additional factors may undermine the effectiveness of
isolation. Mild symptoms or lack of symptoms may ruin the
motivation to respect it, as physical conditions are not an impediment to carry out the daily routine. The legal enforcement of the measure may create tradeoffs discouraging individuals to self-declare as cases~\cite{Lucas2020}. Survey data report that adherence is low~\cite{Smith2021a, Steens2020, Cheng2021}. Among the reported reasons for non-adherence are lower socioeconomic grade, psychological distress, inadequate information and long quarantine duration ~\cite{Brooks2020}.

During COVID-19 pandemic, the duration of isolation has been one
flexible component that authorities adapted from initial estimates of
14 days~\cite{whoisolation} to shorter periods to make the measure more bearable, at the
first signs in summer 2020 showing the difficulty of implementation of
the measure~\cite{Silva2020, Rahmandad2020}. Variable durations mark a tradeoff between a long enough
period of isolation to prevent onward transmission, and a short enough
period that is acceptable by the population. Some countries went as
low as 5 to 7 days to increase adherence~\cite{Germany, France7}, especially in countries where self-isolation was not legally compulsory. Further changes (extension to 10 days~\cite{France10}) occurred later because of the circulation of the Alpha variant (B.1.1.7 lineage),
%the so-called UK variant,
showing the complexity of the biological and social aspects of setting this public health measure~\cite{Ashcroft2021}.

As gaps in any of these aspects may undermine the effectiveness of isolation in aiding epidemic control, here we study through mathematical modeling the role of delayed awareness in entering into isolation and fatigue inducing early release of the measure. 
Our model is a variation of the standard susceptible-infected-susceptible (SIS) compartmental model for infectious disease
dynamics~\cite{PastorSatorras2015, Anderson1991}, allowing for three additional compartments: an isolated (Q)
compartment, an undecided (U) compartment, and a fatigued (F)
compartment.
Here we borrow the classical notation Q commonly used in compartmental models to define the isolation of infectious individuals, notably with the susceptible-infected-quarantined-susceptible (SIQS) model~\cite{Hethcote2002,Chen2020,Zhang2017,EsquivelGomez2018,Young2019, Mancastroppa2020b}. We do not consider the quarantine as the preventive isolation of suspect cases or of contacts of confirmed cases~\cite{Ferretti2020, Lopez2021, Cencetti2020}
, and in the following we will use the terms quarantine and self-isolation as synonyms.
The existence of (temporary) immunity against SARS-CoV-2 would suggest
  the consideration of a SIR-like dynamics. We prefer to consider a SIS-based modeling
  framework as the absence of an immune state allows us to keep analytical derivations simpler
  while still providing (a worse case scenario) intuition on the behavioral mechanisms related to the self-isolation measure. We expect that similar results would be obtained for a SIR-like
model.

The undecided compartment U corresponds to an intermediate state, following infection, during which awareness arises around the knowledge of being infected, involving a delay before the decision to comply with isolation. This state may also be interpreted as the time between infection and testing (thus corresponding to logistical delays in accessing and performing a test, and obtaining the test results), or to the time between infection and symptoms onset (thus corresponding to a pre-symptomatic phase)~\cite{Nature_review, Pullano2020}. Another addition to the standard SIQS model is the possibility that the individual exits self-isolation before its full duration and while still infectious. The compartmental model and transitions are fully explained in the next section. 

We investigate the model on a networked population with a mean-field approach, highlighting how the key parameters describing the epidemic dynamics~--~i.e. the epidemic threshold and the prevalence of infected individuals in the endemic state~--~depend on the different durations associated to these states. We then consider increasingly more accurate mean-field types of approach, allowing to
analyze in detail how individual heterogeneities influence collective properties of the system. We therefore rely on proven effective analytical and numerical tools in order to quantitatively uncover the role of various kinds of imperfections of self-isolation in the spread of a pathogen, which can be
of public health relevance to the control of the currently ongoing pandemic.

\begin{comment}
The paper is structured as follows: in the next section we introduce the epidemic model that takes into account quarantine, delayed awareness and fatigue. We then analyse it from fully homogeneous mean field approach, degree-based Mean Field approach and individual-based mean field approach successively. In the latter two sections we also investigate the effect of the heterogeneity in a couple of chosen parameters: probability of compliance and duration of quarantine.
\end{comment}

\section{Epidemic model with quarantine, delay and fatigue}
%\label{sec:mean-field-theory}

The model we consider is a modification of the usual SIS dynamics~\cite{PastorSatorras2015},
based on the existence of three additional compartments, beyond the standard S (susceptible) and
I (infected) states.

The contact of a susceptible individual with an infectious one leads to a transmission of the
pathogen at rate $\beta$. The newly infected individual enters the state U (undecided)
preceding the decision on whether to self-isolate or not. The decision process is assumed to be
Poissonian with rate $\mu_U$. 
After the decision, the individual enters
the quarantined state Q with probability $p_Q$ (quarantine
probability). With complementary probability he/she instead enters the infected state I.
In the latter case the individual behaves as in the standard infected state of the SIS model.
In the Q state instead the individual has no contacts, and does not transmit
the infection.
Compliance to isolation may however
end before full recovery, as fatigue sets in.
We model this assuming that each quarantined individual transitions to
compartment F (fatigued) at rate $\mu_Q$. Individuals in state F are infectious with the same transmissibility $\beta$
of those in state I.

As the progression of the disease does not depend on the isolation status,
spontaneous recovery transitions occur from states U, I, Q and F to the susceptible state S, at same rate $\mu$. As a consequence of this choice, the average time spent before reaching the S compartment from any of the infected states U, I, Q and F is $1/\mu$, also when there exist multiple paths to recovery. The computation of the average time should indeed receive the contributions of all the possible paths, each weighted by the probability of being chosen; when multiple transitions out of a compartment are possible, the average time of each transition must be conditioned on the fact that the other possible transitions were not undergone.

Our model overall depends on five independent parameters $p_Q$, $\beta$, $\mu_U$, $\mu_Q$
and $\mu$.
\begin{figure}
\includegraphics[width=0.7\columnwidth]{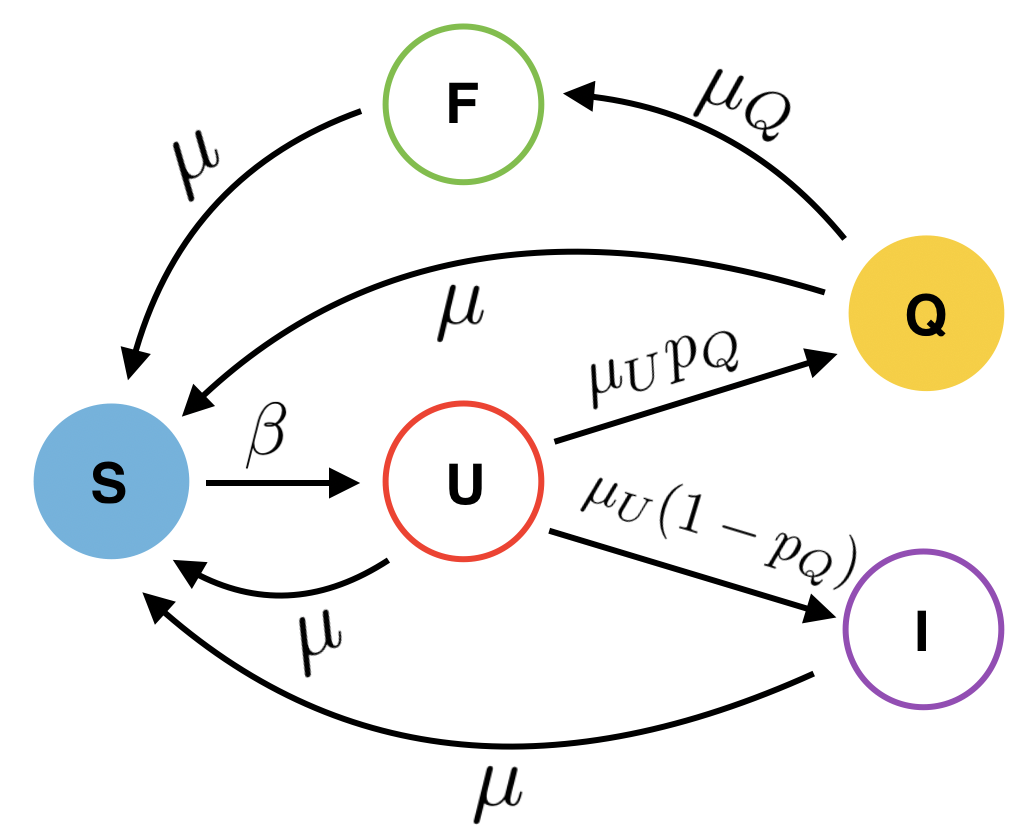}
  \caption{Schematic representation of compartments and transition rates of the model}.
 \label{model}
\end{figure}
The compartments and the transitions allowed between them are depicted in Fig.~\ref{model}.
All transitions are spontaneous except the one taking individuals in state S to the undecided
state U, which occurs because of a contact between a susceptible individual and an
infectious one. Note that in this model individuals in compartments I,
F, U are all infectious with the same transmissibility.

\section{Mean-Field approach}
%\label{sec:quenched-field-theory}

Let us define as $S(t)$, $U(t)$, $I(t)$, $Q(t)$, $F(t)$, the probabilities for an individual
to be in the respective compartments. The sum of these probabilities equals 1,
leaving only 4 independent quantities. We assume a homogeneous pattern
of interactions, with average number of contacts $\av{k}$. Then, the
differential equations describing the evolution of the aforementioned
probabilities read as follows:
\be
\left \{
\begin{array}{ccl}
\dot{U} &=& \beta \av{k} (1-I-U-F-Q) (I+U+F)\\ 
& & -(\mu_U+\mu) U \\
\dot{I} &=& \mu_U  (1-p_Q) U -\mu I\\
\dot{Q} &=& \mu_U  p_Q U -(\mu_Q +\mu) Q\\
\dot{F} &=& \mu_Q Q - \mu F 
\end{array}
\right.
\ee

The disease-free state $(S, U,
I, Q, F) = (1,0,0,0,0)$ is always an equilibrium solution of the system.
Linearization around it shows that it is not stable if
$\lambda=\beta/\mu$ is above the critical value $\lambda_c$, marking
the existence of an endemic state:
\be
\begin{split}
\lambda_c &= \frac{1}{\av{k}} \frac{1}{1-p_Q \frac{\mu \mu_U}{(\mu_Q+\mu)(\mu_U+\mu)}}\\
&=\frac{1}{\av{k}} \frac{1}{1-\frac{p_Q}{\left(1+\frac{T}{T_Q}\right)\left(1+\frac{T_U}{T}\right)}},
\label{lambda_c}
\end{split}
\ee
where the temporal scales $T=1/\mu$, $T_Q=1/\mu_Q$ and $T_U=1/\mu_U$
are the average times spent in the corresponding states.
Eq.~\eqref{lambda_c} contains several known results for limit values of its parameters.
$p_Q=0$ --~individuals never in isolation~-- yields the well-known SIS Mean-field result $\lambda_c=1/\av{k}$. The same limit is
recovered if the quarantine has vanishing duration ($T/T_Q \to \infty$) or when the time to take
a decision diverges ($T_U/T \to \infty$).

For generic values of the parameters Eq.~\eqref{lambda_c} can be
written as
\be
\lambda_c = \frac{1}{\av{k}} \frac{1}{1-p^{eff}_Q}
\ee
where
\be
p^{eff}_Q = \frac{p_Q}{\left(1+\frac{T}{T_Q}\right)\left(1+\frac{T_U}{T}\right)}.
\label{pqeff}
\ee
The quantity $p^{eff}_Q$ is a scaling law turning the effect of
compartments U, F into an effective probability to self-isolate in a
SIQS model.  It is smaller than $p_Q$ and reflects the reduction in
the efficacy of the quarantine due to undecidedness and fatigue.  Note
that, even for full compliance with the quarantine prescription
($p_Q=1$), any value $T_Q<\infty$ or $T_U>0$ is sufficient to make the
threshold finite. For relatively large decision time
  (compared with recovery time) or small quarantine duration, the factor
multiplying $p_Q$ in Eq.~\eqref{pqeff} is small and therefore the
increase of the epidemic threshold for a full quarantine probability
($p_Q = 1$) compared to none ($p_Q = 0$), might be very limited.

In the endemic state, the densities of individuals in the various
compartments are given by
\be
\left \{
\begin{array}{ccl}
U^* &=& \frac{1}{1+T/T_U} \frac{\lambda-\lambda_c}{\lambda}\\
I^* &=& (1-p_Q) \frac{1}{1+T_U/T} \frac{\lambda-\lambda_c}{\lambda}\\
Q^* &=& \frac{p_Q}{(1+T_U/T)(1+T/T_Q)} \frac{\lambda-\lambda_c}{\lambda}\\
F^* &=& \frac{p_Q}{(1+T_U/T)(1+T_Q/T)} \frac{\lambda-\lambda_c}{\lambda}\\
\end{array}
\right.
\ee
where the quantity $\frac{\lambda-\lambda_c}{\lambda}$ is the total density of infected individuals $1-S^* = U^* + I^* + Q^* + F^*$.
The total density of infectious individuals instead is
\be
I^*_{tot} = I^* + U^* + F^* = \frac{1}{\av{k}} \left(\frac{1}{\lambda_c}-\frac{1}{\lambda}\right).
\label{Istar}
\ee

For any $\lambda$, $I_{tot}^*$ depends 
on $p_Q$, $T_Q$ and $T_U$ only via the value of the epidemic threshold.
Equation~\eqref{Istar} then indicates that, for a given $\lambda$, changing parameters in order to increase
the epidemic threshold simultaneously reduces the overall prevalence of infectious
individuals. Therefore maximizing the epidemic threshold, minimizing $I^*_{tot}$ and maximizing $S^* = \lambda_c/\lambda$ are equivalent procedures.

In general, the two parameters $p_Q$ and $T_Q$ are likely to be dependent,
as the perspective of long isolation may discourage people from isolating.
We assume that the quarantine probability is a function $p_Q(T_Q)$
of its duration. In particular we expect that $p_Q$ decreases as $T_Q$ increases.
This leads to the existence of an optimal quarantine duration $T_Q^*$. For
small values of $T_Q$ adherence to quarantine is high, but its duration is
too short, so that people exit from it when they are still infectious.
For large values of $T_Q$ isolated individuals recover when in isolation, but compliance is low. An optimal tradeoff exists between
these two limits.
We want to find the optimal duration of self-isolation $T_Q$ that
minimizes pathogen 
circulation, i.e. it either minimizes $I_{tot}^*$
or maximizes the epidemic threshold.
This occurs when the denominator of the epidemic threshold (Eq.~\eqref{lambda_c}) attains its minimum,
i.e. for $T_Q=T_Q^*$ such that
\be
\left. \frac{dp_Q}{dT_Q}\right|_{T_Q=T_Q^*} = - \frac{p_Q(T_Q^*)}{T_Q^*} \frac{1}{1+\frac{T_Q^*}{T}}.
\label{maximization}
\ee

We make the relationship between $p_Q$ and $T_Q$ explicit, making the
following minimal assumption: 
\be
p_Q(T_Q) = \frac{1}{1 + c T_Q/T},
\label{pQ}
\ee
where the parameter $c$ determines how quickly the probability to enter quarantine decays with its duration.
This way compliance tends to be perfect ($p_Q \to 1$) for extremely short quarantine,
while virtually nobody decides to isolate ($p_Q \to 0$) if the duration of self-isolation is much longer
than the average recovery time.

The optimal quarantine duration, solution of Eq.~\eqref{maximization},
is then
\be
\frac{T_Q^*}{T} = \frac{1}{\sqrt{c}},
\label{T_Qoptimal}
\ee
corresponding to an optimal probability to quarantine $p_Q^*=1/(1+\sqrt{c})$ and to the maximum
threshold
\be
\lambda_c^* = \frac{1}{\av{k}} \frac{1}{1-\frac{1}{\left(1+\sqrt{c}\right)^2\left(1+\frac{T_U}{T}\right)}}.
\ee

\begin{figure}
\includegraphics[width=\columnwidth]{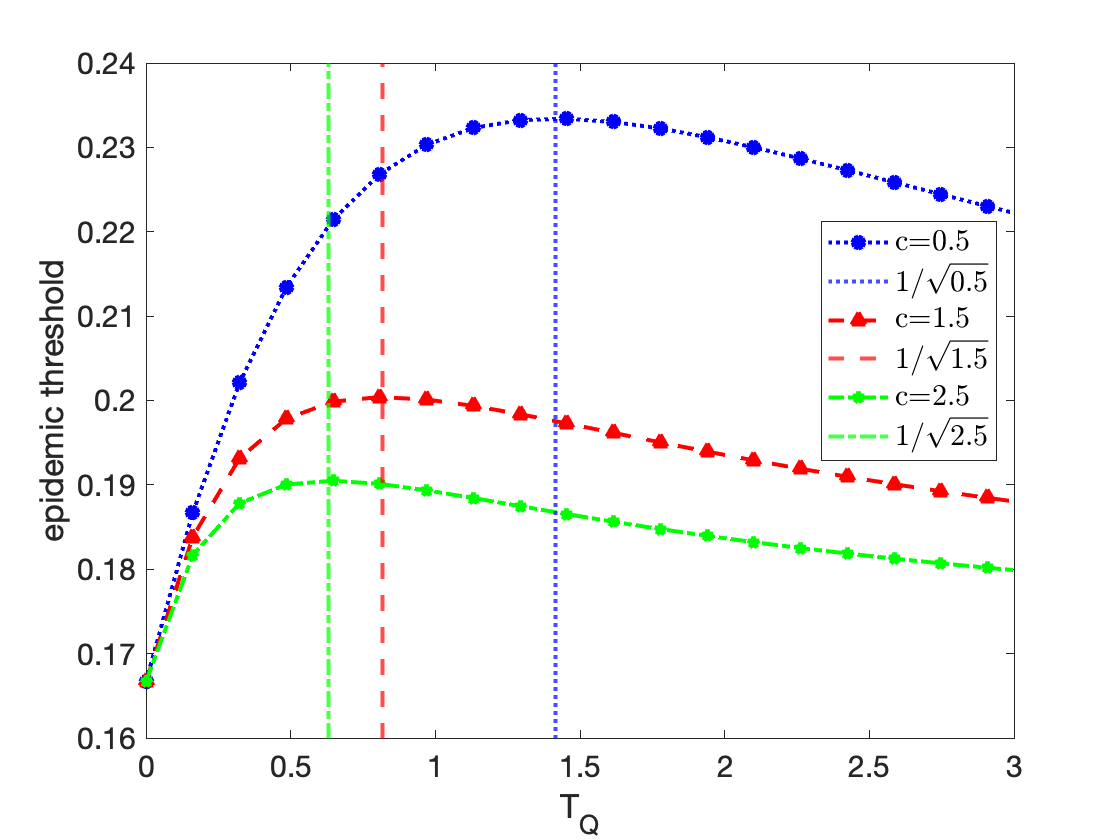}
\caption{Dependence of the epidemic threshold on the quarantine
  duration $T_Q$ for various values of the parameter $c$ in
  Eq.~\eqref{pQ}. Vertical lines indicate the optimal values from
  Eq.~\eqref{T_Qoptimal}. $T=1$, $T_U=0.2$ and $\av{k}=0.6$.}
 \label{optimal}
\end{figure}

We find that for the optimal duration of quarantine $T_Q^*$ the threshold is a decreasing function of the parameter $c$~(see Fig.~\ref{optimal}).

\section{Heterogeneous Mean Field approach}

We now allow for the more realistic assumptions of individuals to have heterogeneous contact rates. We start investigating this
case by means of the Heterogeneous Mean Field (HMF)
approximation~\cite{PastorSatorras2001,PastorSatorras2015},
which assumes that the probability of being in a given compartment only depends
on the degree $k$ of an individual.
Hence the state of the system is described by the set of variables $S_k(t)$, $U_k(t)$, $I_k(t)$, $Q_k(t)$, $F_k(t)$,
where $k$ spans the degree values in the network and normalization to 1 holds for any $k$.
This approach is equivalent to assuming
that the underlying contact network among individuals is annealed~\cite{Dorogovtsev2008}, i.e., connections are fully
rewired at each time step while preserving the degree of each node.
For the standard SIS dynamics on power-law degree-distributed networks with $P(k) \sim k^{-\gamma}$,
the HMF approximation provides very accurate results if $\gamma<5/2$~\cite{Ferreira2012} while
for larger values of $\gamma$ it fails for large systems~\cite{Castellano2010,Ferreira2012}.
For simplicity we further assume that the network is uncorrelated so that
$P(k'|k) = k' P(k')/\av{k}$.

\subsection{Epidemic parameters independent of \texorpdfstring{$k$}{TEXT}}
We first consider the case where all individuals behave in the exact same way so that parameters take fixed values.

The HMF equations are easily written down
\be
\left \{
\begin{array}{ccl}
  \dot U_k &=& \beta (1-I_k-U_k-Q_k-F_k) k \Theta - (\mu_U+\mu) U_k \\
\dot I_k &=& \mu_U  (1-p_Q) U_k -\mu I_k\\
\dot Q_k &=& \mu_U  p_Q U_k -(\mu_Q+\mu) Q_k\\
\dot F_k &=& \mu_Q Q_k - \mu F_k 
\end{array}
\right.
\ee

where $\Theta$ is the probability that a neighbor of a given node is infectious in an uncorrelated network
\be
\Theta = \sum_{k'} \frac{k' P(k')}{\av{k}}(I_{k'}+U_{k'}+F_{k'}).
\label{Theta}
\ee
At stationarity we have
\be
\left \{
\begin{array}{ccl}
0 &=& \beta (1-I_k-U_k-Q_k-F_k) k \Theta - (\mu_U+\mu) U_k \\
0 &=& \mu_U  (1-p_Q) U_k -\mu I_k\\
0 &=& \mu_U  p_Q U_k -(\mu_Q+\mu) Q_k\\
0 &=& \mu_Q Q_k - \mu F_k 
\end{array}
\right.
\ee
whose solution reads
\be
\left \{
\begin{array}{ccl}
  U_k^* &=& \frac{ \lambda k \Theta}{(1 + \mu_U/\mu)(1+\lambda k \Theta)}\\
I_k^* &=& \frac{\mu_U}{\mu} (1-p_Q) U_k^* \\
Q_k^* &=& \frac{\mu_U p_Q}{\mu_Q+\mu} U_k^*\\
F_k^* &=& \frac{\mu_Q}{\mu} \frac{\mu_U p_Q}{\mu_Q+\mu} U_k^*
\end{array}
\right.
\label{stationary_density}
\ee

Inserting the stationary values into Eq.~\eqref{Theta} we find
\be
\begin{array}{ccl}
\Theta = \sum_{k'} \frac{k' P(k')}{\av{k}} \left[1-\frac{p_Q}{(1+\mu_Q/\mu)(1+\mu/\mu_U)}\right](1 + \mu_U/\mu) U^*_{k'}
\end{array}
\label{Theta_2}
\ee A nontrivial solution $\Theta>0$ only exists if
  the derivative with respect to $\Theta$ of the r.h.s. of
  Eq.~\eqref{Theta_2} [where we substitute $U_{k'}^{*}$ by its
    explicit dependence on $\Theta$ using
    Eq.~\eqref{stationary_density}] evaluated for $\Theta=0$, is
  larger than 1. This condition allows us to determine the epidemic
threshold \be \lambda_c =
\frac{1}{1-\frac{p_Q}{\left(1+\frac{T}{T_Q}\right)\left(1+\frac{T_U}{T}\right)}}
\frac{\av{k}}{\av{k^2}}.
\label{lambda_cHMF}
\ee
We observe that this threshold is simply the
HMF threshold for SIS~\cite{PastorSatorras2001} modulated by a factor that takes into account quarantine probability, duration as well as delay. The effect of
topology factorizes. For a homogeneous network
Eq.~\eqref{lambda_c} is recovered, since $\av{k^2}=\av{k}^2$.

We perform numerical checks of these predictions, by simulating
SIS dynamics (using a Gillespie optimized algorithm~\cite{Cota2017})
on networks built according to the uncorrelated configuration
model~\cite{Catanzaro2005}. In this model, we consider an upper cutoff on the degrees --~$k \in [k_{min} = 3, k_{max} = \sqrt{N}]$~-- in order to have an uncorrelated network without multiple and self connections.
The epidemic threshold is estimated by finding the value of $\lambda=\beta/\mu$ at which the susceptibility of the system reaches a maximum value ~\cite{Ferreira2012}. Such susceptibility is computed for the number of infected individuals in the quasistationary regime (the order parameter of the epidemic phase transition). Of course only surviving runs of the dynamics need to be considered. In order to work with the equivalent of surviving runs, we implemented the so-called Quasistationary State method (QS) ~\cite{Ferreira2012}, for which the dynamics never allows the system to enter the healthy absorbing state.

In the following, we consider networks with an exponent $\gamma=2.25$ of the degree distribution and, unless otherwise specified, with a network size $N=10^5$.
%\textcolor{red}{In the range of $\gamma<3$ we expect that, just like for the SIS model, HMF, although in general less accurate, accidentally outperforms QMF. The qualitative inaccuracy of HMF would only become evident for $\gamma>3$~\cite{Boguna2013}.}

\begin{figure}
\includegraphics[width=\columnwidth]{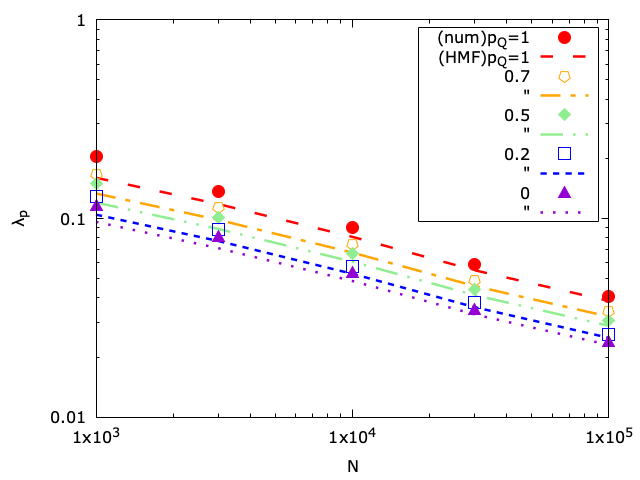}
\caption{
  Epidemic threshold $\lambda_c$ as a function of
  the system size $N$ for $\gamma=2.25$ and several values of $p_Q$.
  Dashed lines represent the theoretical prediction,
  Eq~\eqref{lambda_cHMF}, while symbols are the numerical estimates,
  obtained from the peak of the susceptibility~\cite{Cota2017}. $T=1$, $T_U=0.25$, $T_Q=1$.}
 \label{lambdacvsN}
\end{figure}

In Fig.~\ref{lambdacvsN} we plot the epidemic threshold as a function
of the system size $N$ for several values of the probability $p_Q$ to
enter quarantine.  We first note an excellent agreement between the
theory (dashed lines) and the simulations (symbols), further
increasing as $N$ grows.  The value of the threshold decreases as a
function of size, as a consequence of the diverging second moment at
the denominator of Eq.~\eqref{lambda_cHMF}.
%\textcolor{red}{We note
%  that it is only for $\gamma<3$ that we expect HMF to provide such a
%  good qualitative behavior with system size, as the second moment of
%  the distribution only diverges, and hence correctly predicts a
%  vanishing epidemic threshold, in that range of $\gamma$.}
A higher
quarantine probability leads to an increase of the threshold but for the present choice of
  parameter values ($T = 1$, $T_U=0.25$, $T_Q=1$),
  the effect is not dramatic: even a complete
participation to quarantine ($p_Q=1$) implies only a (slightly more
than) two-fold increase in the value of the threshold with respect to
the $p_Q=0$ case.
\begin{figure}
\includegraphics[width=\columnwidth]{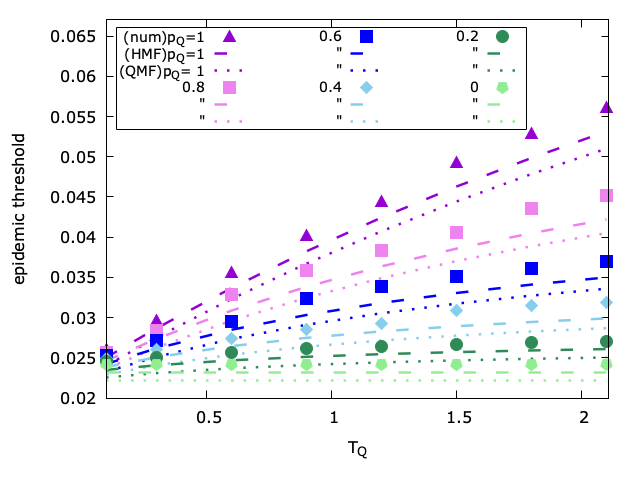}
\caption{Epidemic threshold $\lambda_c$ as a function of $T_Q$ for $T=1$, $T_U=0.2$ and various $p_Q$.
 Symbols are the results of numerical
  simulations, dashed lines are the predictions of HMF theory, dotted lines are the predictions
of QMF theory.}
\label{lambdacvsTQ1}
\end{figure}
In Fig.~\ref{lambdacvsTQ1} we show the dependence of the threshold
on the duration of quarantine for various values of the probability $p_Q$.
We observe that a longer duration of quarantine
leads to a larger epidemic threshold, but the effect is sizeable
only provided $p_Q$ is quite large.
\begin{figure}
\includegraphics[width=\columnwidth]{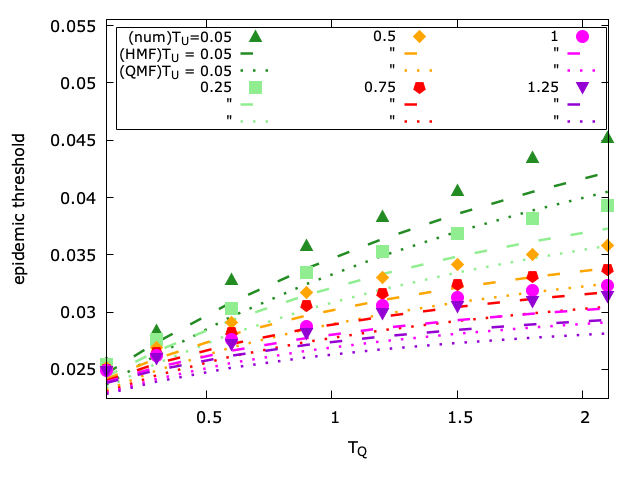}
\caption{Epidemic threshold $\lambda_c$ as a function of $T_Q$ for $T=1$, $p_Q=0.7$ and various $T_U$.
 Dashed lines are the predictions of HMF theory and dotted lines are the predictions of QMF theory while symbols are the results of numerical simulations. }
\label{lambdacvsTQ2}
\end{figure}
In Fig.~\ref{lambdacvsTQ2} we show the dependence of the threshold
on the duration of quarantine for various delays $T_U$.
Here too the threshold increases smoothly with $T_Q$.
If the decision time is much smaller than the time to heal the effect becomes
relevant also for reasonable values of $T_Q$.

\begin{figure}
\includegraphics[width=\columnwidth]{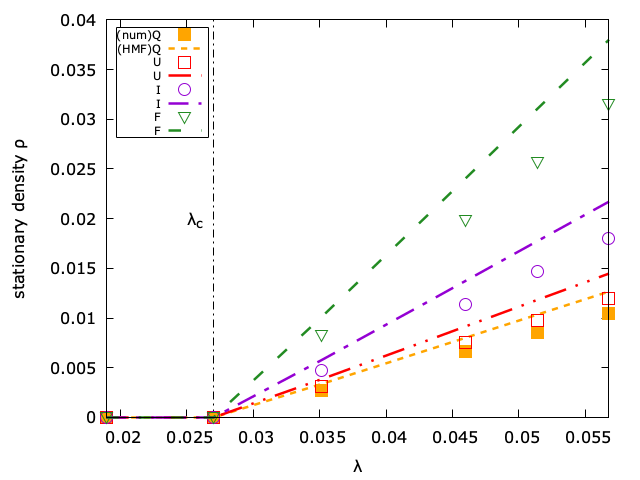}
\caption{Density of individuals in the various compartments in the stationary state as a
  function of $\lambda$ for $T=1$, $T_U=0.2$, $T_Q=0.33$ and $p_Q=0.7$.}
\label{rhovslambda}
\end{figure}

Finally in Fig.~\ref{rhovslambda} we report the dependence of the total density of
individuals in the various compartments as a function of $\lambda$, showing a fair
agreement between theory and numerics. As long as the parameters of the model are finite and strictly positive, each of these densities carries information on the global state of the epidemic, each of them simply being a fraction of the order parameter.

\subsection{Degree-dependent epidemic parameters}
Empirical evidence~\cite{Muscillo2020} suggests that people having more contacts
or being more active tend to be more reluctant in reducing their interactions
to prevent contagion. This may be due to the fact that each interrupted contact carries with it a substantial economical, social and/or psychological cost. Real data also suggest that individuals with high activity are more attractive, which may make it more difficult for them to self-isolate given the numerous solicitations they receive from others~\cite{Mancastroppa2020b}.
It is then quite natural to believe that also adherence to the
prescription to self-isolate may be different (and in particular be suppressed) for people with a large
number of contacts. 
In our framework it is possible to model such a realistic element
by assuming that the probability to enter quarantine and/or its duration depend
on the degree of the node, a proxy of individual activity.
In particular it is reasonable to expect both $p_Q$ and $T_Q$ to decrease with $k$.

\begin{figure}
\includegraphics[width=\columnwidth]{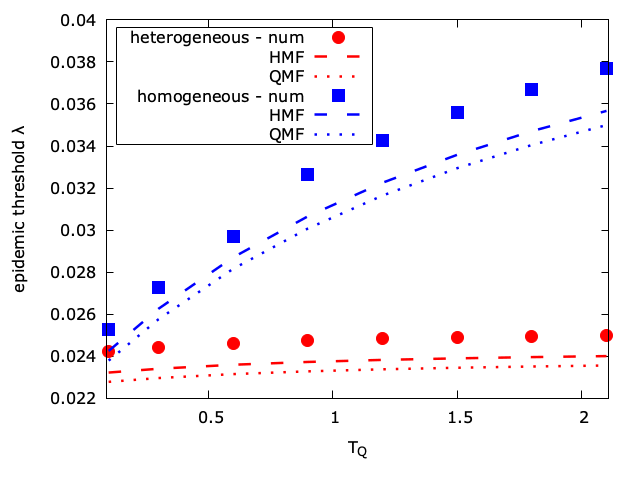}
\caption{Epidemic threshold as a function of $T_Q$ for $T=1$, $T_U=0.25$. In the heterogeneous
  case (circles and bottom lines) $p_Q(k)=k_{min}/k$. In the homogeneous case (squares and top lines)
  $p_Q$ is the same for all nodes: $p_Q=\av{p_Q(k)}=0.66$. Dashed lines are for the HMF predictions whereas dotted lines are for the QMF ones.}
\label{pQdep}
\end{figure}

By repeating the calculations already performed in the case with degree-independent parameters
we easily find that the epidemic threshold reads
\be
\lambda_c = \frac{\av{k}}{\av{\left(1-\frac{p_Q(k)}{\left(1+\frac{T}{T_Q(k)}\right)\left(1+\frac{T_U}{T}\right)} \right)k^2}},
\ee
where $\langle X(k) \rangle = \sum_k P(k) X(k)$.
Hence, depending on how $p_Q$ and $T_Q$ behave for large $k$, quarantine may
reduce or not the vulnerability of scale-free networks to epidemics.
We test this prediction again by performing simulations on networks built
according to the uncorrelated configuration model. For reference, we
compare with results obtained with degree-independent parameters tuned
to have exactly the same average value of the degree-dependent case.
We first check what happens assuming $p_Q(k) = k_{min}/k$, so that
compliance is perfect for nodes having minimal connectivity while
it becomes very small for large $k$.
In Fig.~\ref{pQdep} we report the behavior of the epidemic threshold as
a function of $T_Q$ for the degree-dependent case and for a degree independent
case such that $p_Q = \av{p_Q(k)}$. 
It turns out that the threshold is smaller in the degree-dependent case and in
particular that it grows much more slowly with $T_Q$. The effect of a long self-isolation
of less connected individuals is almost completely offset by the little compliance
of nodes of large degree.

\vspace{1cm}

\begin{figure}
\includegraphics[width=\columnwidth]{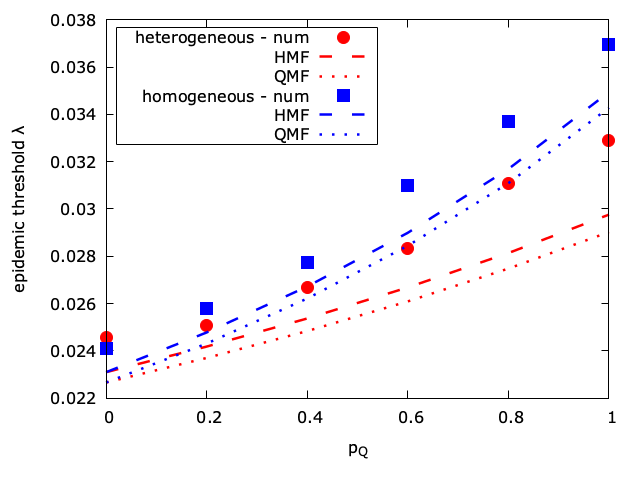}
  \caption{Epidemic threshold as a function of $p_Q$ for $T=1$, $T_U=0.25$. In the heterogeneous
  case (circles and bottom lines) $T_Q(k)=(T-T_U)(k_{max} - k)/(k_{max} - k_{min})$. In the homogeneous case (squares and top lines)
  $T_Q$ is the same for all nodes $T_Q=\av{T_Q(k)}=0.71$. Dashed lines are for the HMF predictions whereas dotted lines are for the QMF ones.}.
\label{TQdep}
\end{figure}

We then check what happens instead when $T_Q(k) = (T-T_U)(k_{max} - k)/(k_{max} - k_{min})$. Complying individuals with few contacts self-isolate until full recovery, whereas individuals with large connectivities spend a vanishing time in quarantine. This degree-dependent scenario is compared in Fig.~\ref{TQdep} to the degree-independent one in such a way that in the latter $T_Q = \av{T_Q(k)}$. We find that the possibility for a few hubs to undergo shorter isolation periods than the average individual of the population lowers the epidemic threshold the more the larger the quarantine probability. The linear interpolation between the extreme behaviours of hubs and poorly-connected individuals seems to be slow enough not to completely offset the advantages put forward by the self-isolation prescription. 

\section{Quenched Mean Field approach}

\subsection{Epidemic parameters independent of \texorpdfstring{$k$}{TEXT}}

A more refined approach to the model dynamics on networks is provided by the Quenched Mean Field
(QMF) approximation~\cite{Wang2003,VanMieghem2009,Gomez2010} (also known as Individual Based mean-field approach),
which takes into account the detailed structure
of the network encoded in the adjacency matrix $A_{ij}$. Defining as $I_i(t)$, $U_i(t)$, $Q_i(t)$, $F_i(t)$
the probabilities that node $i$ is in state I, U, Q and F respectively, the evolution of the model
is described by the set of $4N$ equations
\be
\left \{
\begin{array}{ccl}
  \dot U_i &=& \lambda (1-I_i-U_i-Q_i-F_i) \cdot \\
  & & \cdot \sum_{j} A_{ij} (I_j+U_j+F_j) -(\mu+\mu_U) U_i \\
\dot I_i &=& \mu_U (1-p_Q) U_i -\mu I_i\\
\dot Q_i &=& \mu_U p_Q U_i -(\mu+\mu_Q) Q_i\\
\dot F_i &=& \mu_Q Q_i - \mu F_i 
\end{array}
\right.
\ee

By linearizing around the healthy state and imposing the largest eigenvalue of the Jacobian matrix
to be equal to 0 one obtains the epidemic threshold
\be
\lambda_c = \frac{1}{1-\frac{p_Q}{\left(1+\frac{T}{T_Q}\right)\left(1+\frac{T_U}{T}\right)}}
\frac{1}{\rho(A)},
\label{lambdac3}
\ee where $\rho(A)$ is the spectral radius of the adjacency matrix
(i.e. its largest eigenvalue).  This expression is perfectly analogous
to the QMF result for standard SIS dynamics (corresponding to
Eq.~\eqref{lambdac3} for $p_Q=0$), for which it is well
known~\cite{Goltsev2012,PastorSatorras2015} that the QMF threshold is
a lower bound of the true threshold.  We expect this to be true also
for the present modification of the SIS model. Indeed
Figs.~\ref{lambdacvsTQ1} and~\ref{lambdacvsTQ2} show that
Eq.~\eqref{lambdac3} is a tight lower bound, as it is the case for
$\gamma=2.25<5/2$.  For larger values of $\gamma$ instead additional
nontrivial effects~\cite{Castellano2020} make the QMF estimate
inaccurate for large networks. We note also that the
  HMF predictions are slightly closer than QMF to the numerical
  results. This better performance of HMF with respect to
  QMF (which occurs also for SIS~\cite{Boguna2013}) is accidental: the additional
  approximation introduced by HMF partly cancels the error due to the QMF approximation.
  It is only for larger values of $\gamma>3$ and larger system sizes that QMF reveals
  its more accurate qualitative behavior.

\subsection{Degree-dependent epidemic parameters}

If the parameter $p_Q$ depends on $k$ we obtain that the threshold is
\be
\lambda_c = \frac{1}{\rho({\tilde A})}.
\ee
The quantity $\tilde A$ is a modification of the adjacency matrix of the system
\be
{\tilde A} = A \left(\mathbb{1}-\mathbb{P}_Q \frac{1}{\left(1+\frac{T}{T_Q}\right)\left(1+\frac{T_U}{T}\right)}\right),
\ee
where $\mathbb{P}_Q=diag(p_Q(k(i)))$.

If instead the parameter $T_Q$ depends on $k$ we obtain that the threshold is
\be
\lambda_c = \frac{1}{\rho({\tilde A})}.
\ee
$\tilde A$ is now
\be
   {\tilde A} = A \left(\mathbb{1}-p_Q \frac{1}{\left(\mathbb{1}+\frac{T}{\mathbb{T_Q}}\right)\left(1+
     \frac{T_U}{T}\right)}\right),
\ee
where $\mathbb{T}_Q=diag(T_Q(k(i))))$.

\section{Conclusions}

In this paper we have defined and used an epidemic model to study the
role of delayed case detection/infection awareness, compliance to
self-isolation, and fatigue-induced early drop-out on the
effectiveness of self-isolation as a non-pharmaceutical
intervention. We found that adherence to the prescription to
self-isolate once infected scaled up the epidemic threshold compared
to the simple SIS result, and delay into entering isolation or early
release from it resulted in a reduction of effective adherence to
self-isolation. If the propensity to enter self-isolation and the time
spent isolated decrease with the individual number of contacts
(degree), the low adherence of few well socially connected individuals
may undermine the effectiveness of the entire non-pharmaceutical
measure against the epidemic. This is a key result as it suggests that
these phenomena, empirically found~\cite{Lauer2020, Smith2021a}, may
strongly limit the impact of isolation programs on the pandemic,
unless specific measures are implemented to overcome these
barriers~\cite{Smith2021a}.

The applicability of this model to real case scenarios would take advantage from being informed by real data for both the parameters describing the COVID-19 dynamics and behavioral parameters. Whereas the former can be obtained by fitting case incidence time-series, the latter rely on a complex interaction between top-down regulations and behavioral adaptations and are hence harder to be inferred from data. The behavioral parameters are therefore explored rather than fit to data. We plan to include data from surveys in a future development of this study. Also additional ingredients must be considered in the model to increase the applicability of the model
to more realistic scenarios. First, a more detailed compartmental
structure accounting for the different phases of COVID-19 disease
progression, to better account for the interplay of different time
periods and include asymptomatic and paucisymptomatic states. These
may also result in different behaviors, reducing adherence and
increasing early drop-outs, compared to symptomatic cases. Our
simplified approach has, however, the advantage of being analytically
tractable, therefore providing an immediate solution under certain
approximation and offering an intuition into the behavior of the
system. In this perspective, the SIS model was preferred to a model
with immunity. Second, transitions were modeled with Poissonian
probability distributions, whereas many of these processes are
generally described by broader distributions~\cite{Bonaccorsi2016,
  deArruda2020} . In this context, we expect that this approximation
may impact the recovery process from the state F differently than from
the state I.

Further directions can be considered to expand this approach in future
work. Here, we assumed that isolation prevents all contacts. In
reality, isolation is never 100\% effective, due both to behavioral
aspects, and hard living constraints (e.g., household crowding~\cite{Valdano2021}).
Different degrees of imperfect isolation can be considered in terms of
approximations altering the contact pattern. We did not consider in
this study the role of quarantine as preventive isolation of suspect
cases or contacts of confirmed cases. There is now a large body of
literature on the role of contact tracing in combination to isolation
and testing in COVID-19 control~\cite{Ashcroft2021}, and the importance of speeding up
this process through digital tools~\cite{Ferretti2020}. Beside contact tracing, the
introduction of a compartment describing individuals uncertain about
their infection status, but still with recommended self-isolation, constitutes an additional component in limiting adherence, as motivation to self-isolate is reduced in absence of symptoms or of a test result confirmation. These processes are likely to be governed by different parameters of quarantine probability and duration. Finally, adherence to self-isolation may be the result of an individual component, explored here, along with a population component defined by a level of awareness and of risk perception that may evolve over time~\cite{Smith2020b}, depending on the evolving epidemic context. This may be an important component contributing to the observed relaxation effects after COVID-19 pandemic wave, possibly resulting in case resurgences~\cite{Amaral2021, Fair2021, Johnston2020}.

\bibliography{CCVDM1.bib}

%\appendix

\end{document}